\documentclass[letterpaper,12pt]{article}

\usepackage[top=2.0cm,bottom=2.8cm,left=3.0cm,right=3.0cm]{geometry}
\usepackage[authoryear]{natbib}
\usepackage{graphicx}
\usepackage{amssymb}
\usepackage{multicol}
\usepackage{authblk}
\usepackage[usenames,dvipsnames]{xcolor}
\usepackage{listings}
\usepackage{setspace}

\usepackage{parskip}
\usepackage[margin=7pt,font=small,labelfont=bf,labelsep=quad]{caption}

\definecolor{light-gray}{gray}{0.95}

\bibpunct{(}{)}{;}{a}{}{,}

\newcommand{\farcsec}{\mbox{\ensuremath{.\!\!^{\prime\prime}}}}

\begin{document}

\begin{center}

{\bf \LARGE Go deep, not wide\footnote{
A white paper submitted in response to the National Radio Astronomy Observatory's\linebreak
{\it Call for White Papers on the VLA Sky Survey (VLASS)} released 2013 September 9
(https://science.nrao.edu/science/surveys/vlass)}}
\vspace{5mm}

Christopher A. Hales\footnote{Jansky Fellow;
National Radio Astronomy Observatory, P.O. Box 0, Socorro, NM 87801, USA}
\vspace{12mm}

\begin{quote}
\begin{center}{\bf \large Abstract}\end{center}
The Karl G. Jansky Very Large Array (VLA) is currently the world's most powerful cm-wavelength
telescope. However, within a few years this blanket statement will no longer be entirely true,
due to the emergence of a new breed of pre-SKA radio telescopes with improved surveying
capabilities. This white paper explores a region of sensitivity-area parameter space where an
investment of a few thousand hours through a VLA Sky Survey (VLASS) will yield a unique dataset
with extensive scientific utility and legacy value well into the SKA era: a deep full-polarization
L-band survey covering a few square degrees in A-configuration. Science that can be addressed with
a deep VLASS includes galaxy evolution, dark energy and dark matter using radio weak
lensing, and cosmic magnetism. A deep VLASS performed in a field with extensive multiwavelength
data would also deliver a gold standard multiwavelength catalog to inform wider and shallower
surveys such as SKA1-survey.
\end{quote}

\end{center}
\vspace{10mm}

\vspace{4mm}
\noindent{\bf \large Introduction}

\vspace{1mm}\noindent If a sky survey is to be performed with the VLA, displacing highly
competitive PI-led science for a few thousand hours, it will need to facilitate
a wide range of scientific goals and have lasting value for many years. The VLA has 3 options:
go as wide as possible \citep[e.g. NVSS;][]{1998AJ....115.1693C}, as deep as possible
\citep[e.g.][]{2008AJ....136.1889O}, or somewhere in the middle \citep[e.g. FIRST, the VLA
Galactic plane survey, or Stripe 82;][]{1995ApJ...450..559B,2006AJ....132.1158S,2011AJ....142....3H}.

If the VLASS adopts a medium/wide-field approach, it will face significant competition from
3 key facilities in the near future. These are the Apertif focal-plane upgrade on the
Westerbork Synthesis Radio Telescope \citep[WSRT;][]{2009wska.confE..70O}, and the MeerKAT
\citep{2009arXiv0910.2935B} and ASKAP \citep{2008ExA....22..151J} SKA-pathfinders.

To avoid overlap with these upcoming facilities, this white paper explores parameter space
for a deep field where the VLA can capitalize on its key strengths of sensitivity and angular
resolution to provide the astronomical community with a unique survey.

\newpage
\noindent{\bf \large Identifying unique parameter space for a VLA sky survey}

\begin{figure}
 \centerline{\includegraphics[clip, angle=-90, totalheight=130mm]{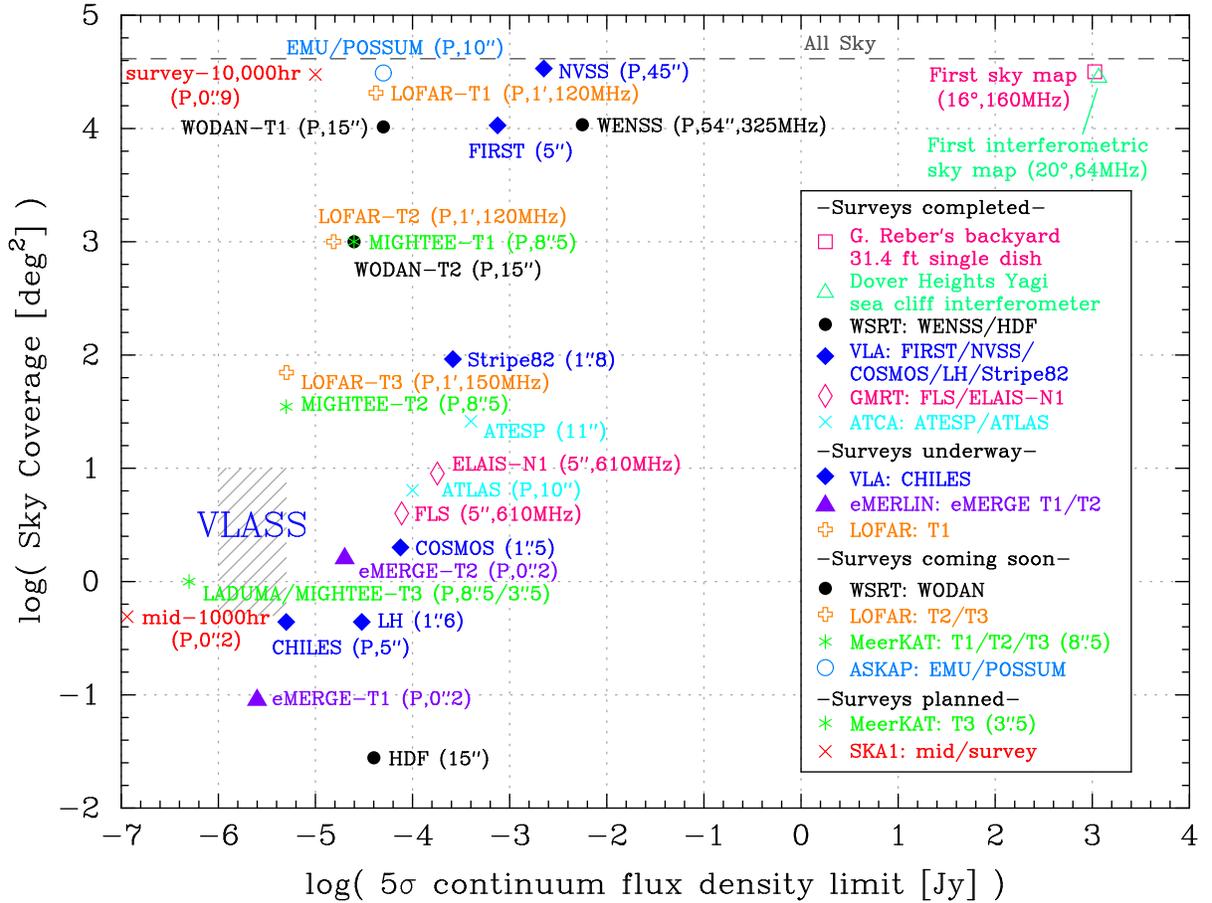}}\vspace{5mm}
 \caption[caption]{
Comparison of selected radio sky surveys, scaled to 1.4~GHz assuming a spectral index of -0.8.
Different telescopes are coded by symbol/color as indicated by the legend. The legend also indicates
which surveys are completed, ongoing, coming soon, or planned for future telescopes. The terms in
parentheses specify if a polarization catalog was produced (P), the angular resolution, and the
observing frequency if not at 1.4~GHz. For a given amount of telescope time, a survey must trade
sensitivity with sky coverage; deep and wide surveys (upper left corner) are the most expensive.
A simplified version of this figure is presented in Fig.~\ref{fig:surveys2}.
\hspace*{\fill}\vspace{3mm}\endgraf{\scriptsize
References: first sky map \citep{1944ApJ...100..279R}, first interferometric sky map \citep{1946Natur.157..296H},
WSRT-WENSS \citep{1997A&AS..124..259R,2007A&A...461..963S}, WSRT-HDF \citep{2000A&A...361L..41G},
WSRT-WODAN \citep{2011JApA...32..557R}, VLA-FIRST \citep{1995ApJ...450..559B}, VLA-NVSS \citep{1998AJ....115.1693C},
VLA-COSMOS \citep{2007ApJS..172...46S}, VLA-LH \citep{2008AJ....136.1889O}, VLA-Stripe82 \citep{2011AJ....142....3H},
VLA-CHILES (COSMOS HI Large Extragalactic Survey; 1000~hr single pointing, started 2013 Oct),
GMRT-FLS \citep{2007MNRAS.376.1251G}, GMRT-ELAIS-N1 \citep{2008MNRAS.383...75G}, ATCA-ATESP \citep{2000A&AS..146...41P},
ATCA-ATLAS \citep{hales}, eMERLIN-eMERGE \citep{emerge}, LOFAR \citep{lofar}, MeerKAT-LADUMA/MIGHTEE
\citep{2012IAUS..284..496H,2012AfrSk..16...44J}, ASKAP-EMU/POSSUM \citep{2010AAS...21547013G,2011PASA...28..215N},
SKA1-mid/survey \citep{ska1}. Apologies if your favorite facility/survey is not included.\endgraf}
 }
 \label{fig:surveys}
\end{figure}

\vspace{1mm}\noindent Fig.~\ref{fig:surveys} displays a sample of existing, upcoming,
planned, and future surveys with various telescopes. It is clear that within a few years,
WSRT, MeerKAT, and ASKAP will have begun and perhaps completed all-sky surveys at L-band
in full polarization with angular resolution $>8^{\prime\prime}$, good sensitivity to
extended structures, and well-suited to the study of radio transients. Instead of
performing a medium/wide-field survey of its own,\linebreak the VLA may be better suited to
targeted follow-up of interesting sources utilizing flexible observing modes.

There is, however, a region of parameter space in Fig.~\ref{fig:surveys} (shaded)
that will not be explored until the SKA-era, and which is well-suited for a VLASS:
a deep-field survey covering a few square degrees with arcsecond resolution at L-band.
Science goals that can be pursued with such a deep-field VLASS are presented later in
this white paper.

\begin{figure}
 \centerline{\includegraphics[clip, angle=-90, totalheight=144mm]{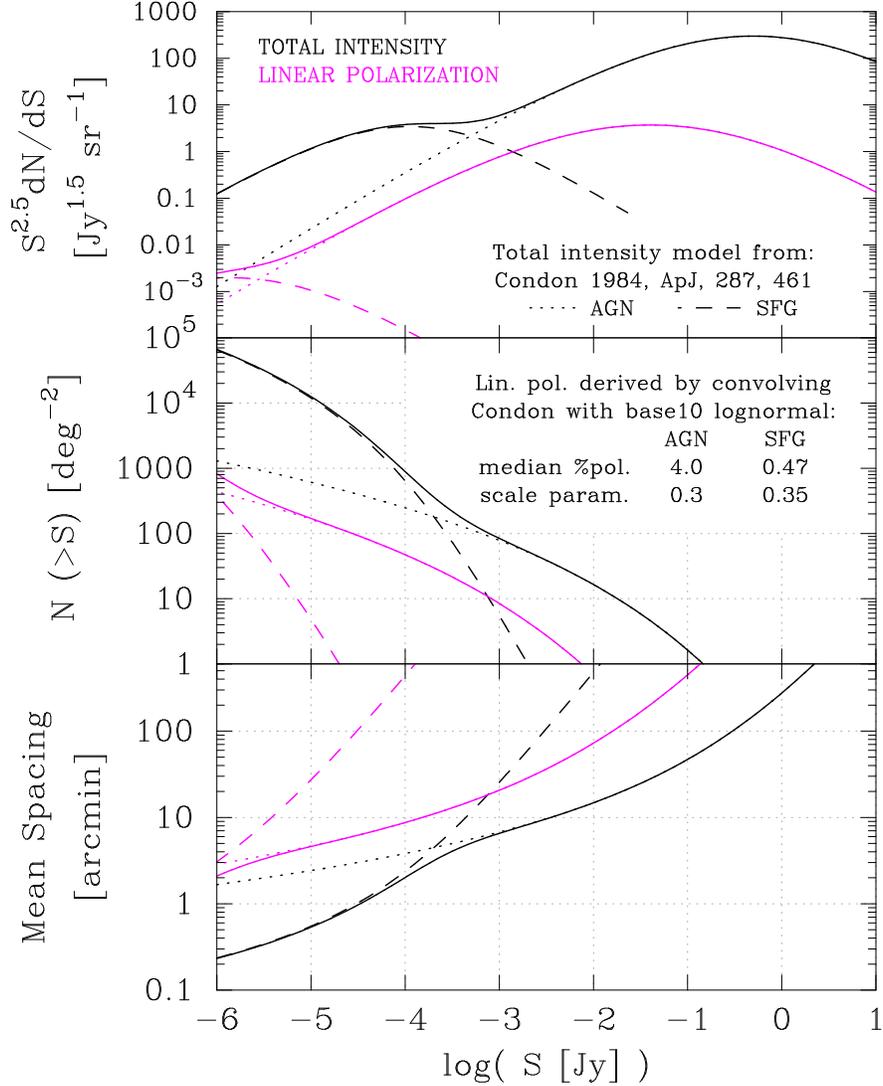}}\vspace{5mm}
 \caption{
Estimated properties of the 1.4~GHz extragalactic sky down to 1~$\mu$Jy. Shown are total intensity
and linear polarization Euclidean normalized differential source counts (top panel), integral counts
(middle panel), and mean spacing between sources (lower panel). Total intensity estimates are from
the model by \citet{1984ApJ...287..461C} for sources powered primarily by active galactic nuclei
(dotted curves) and star formation (dashed curves); this model fits modern data well
\citep[e.g.][]{2012ApJ...758...23C}. Linear polarization estimates were obtained for each source
class by convolving the total intensity differential counts by a distribution of fractional
polarization. The latter were modeled with the lognormal form
$[2 \pi x \ln(10)\sigma]^{-0.5} \exp\{ -[\log_{10}(x/m)]^2 / (2\sigma^2)\}$, with
$m$ and $\sigma$ given respectively by the median fractional polarizations and scale parameters
indicated in the middle panel. The values of $m$ and $\sigma$ for active galaxies are from
\citet{hales}, while those for star forming galaxies are estimated from \citet{2009ApJ...693.1392S}.
There is observational evidence supporting the total intensity model down to 40~$\mu$Jy (direct: counts)
and 2~$\mu$Jy (indirect: confusion analysis), and the linear polarization model down to 100~$\mu$Jy
(direct: counts).
 }\label{fig:counts}
\end{figure}

As with FIRST and NVSS, the optimal VLA frequency band to maximize the sky
density of extragalactic sources is L-band. This band is also particularly
attractive for HI and polarization science (both key drivers for SKA-pathfinder
telescopes and the SKA itself), and because the VLA can achieve $<2^{\prime\prime}$
angular resolution in A-configuration (this point is expanded on below). This white
paper therefore adopts L-band as optimal. For reference, Fig.~\ref{fig:counts} presents
estimated source counts in total intensity and linear polarization for the L-band
extragalactic sky.

\begin{figure}
 \centerline{\includegraphics[clip, angle=-90, totalheight=100mm]{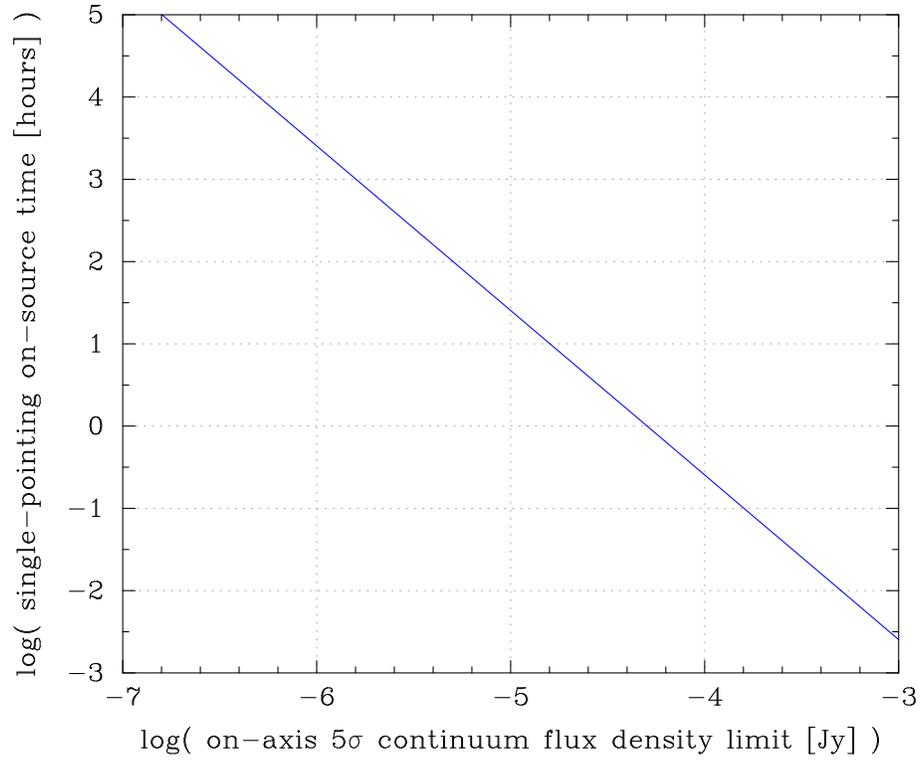}}\vspace{5mm}
 \caption{
Time on-source (no calibration overheads included) required to reach a given on-axis $5\sigma$
detection threshold for a single-pointing observation in A-configuration, assuming 600~MHz RFI-free bandwidth
at 1.5~GHz and natural weighting.
 }\label{fig:time}
\end{figure}

\begin{figure}
 \centerline{\includegraphics[clip, angle=-90, totalheight=90mm]{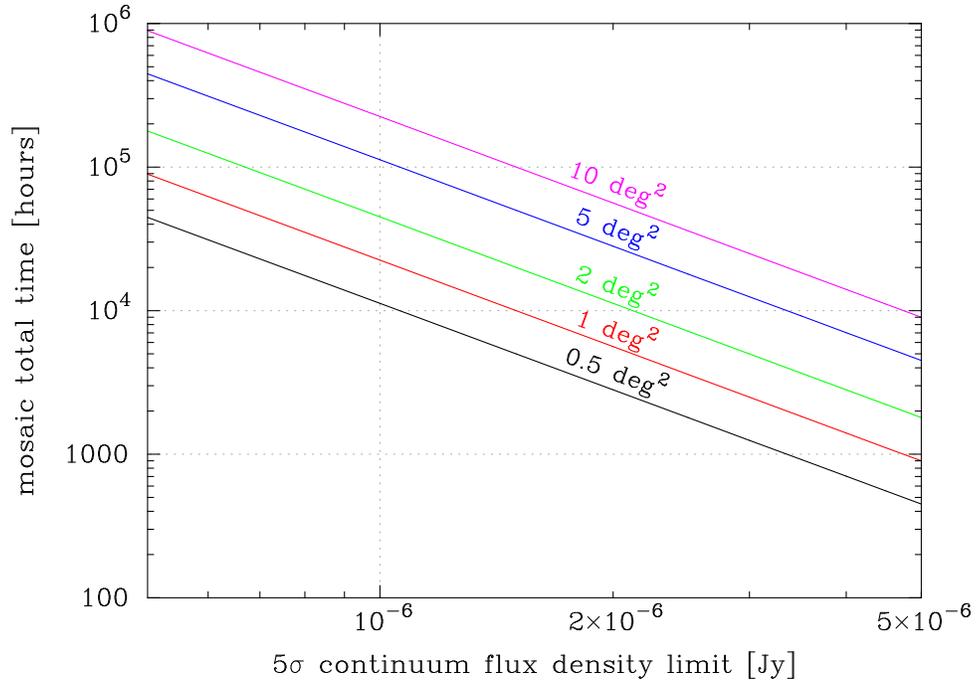}}\vspace{5mm}
 \caption{
Total observing time (including 25\% calibration overheads) required to produce a mosaic with uniform
sensitivity over a given area in A-configuration, assuming 600~MHz RFI-free bandwidth at 1.5~GHz and natural weighting.
Total observing time is calculated as $2.2\,tA/\theta^2$, where $t$ is the on-source time require to reach a
given on-axis sensitivity in a single pointing, $A$ is the mosaic area, and $\theta$ is the primary beam FWHM.
 }\label{fig:mosaic}
\end{figure}

The boundaries for the shaded parameter space in Fig.~\ref{fig:surveys} are selected
as follows. The $5\sigma$ flux density upper boundary is set to 5~$\mu$Jy in order
to probe deeper than the COSMOS HI Large Extragalactic Survey (CHILES; recently
started taking data with the VLA), and to ideally obtain more than a few polarized
starburst galaxies per square degree (see top panel of Fig.~\ref{fig:counts}).
The survey area upper boundary is set to 10~deg$^2$ to keep the total time request
below 10,000~hours for a $5\sigma$ flux density upper bound of 5~$\mu$Jy.
For reference, Fig.~\ref{fig:time} and Fig.~\ref{fig:mosaic} present time estimates
to reach various $5\sigma$ flux density detection thresholds for a range of survey
sizes. The $5\sigma$ flux density lower boundary is set to 1~$\mu$Jy to ensure that a
survey area lower boundary of 0.5~deg$^2$ can be observed within a total time
request of 10,000~hours. Survey areas $<0.5$~deg$^2$ can be observed within a
single pointing at L-band, though areas $>1$~deg$^2$ are likely be preferred
for a deep VLASS to minimize influence from cosmic or sample variance.

A-configuration is likely optimal for a deep VLASS for the following reasons.
First, arcsecond angular resolution is required for morphological studies of
faint radio galaxies \citep{2005MNRAS.358.1159M}; VLA A-configuration may be
better suited for such studies than the MeerKAT LADUMA/MIGHTEE tier-3 survey with
$3\farcsec5$ resolution. Second, the effects of radio interference at
L-band are minimized in A-configuration, where the antennas are as far apart as possible.
Third, for a given total observing time, the percentage of time lost from PI-led science
to a VLASS in A-configuration will be less than for any other configuration (particularly
hybrids). Fourth, if a deep\linebreak A-configuration VLASS is performed over a target field with
eMERLIN-eMERGE data (field selection is discussed later in this white paper), then the
two surveys can be combined to provide sensitivity to angular scales from
$40^{\prime\prime}-0\farcsec2$. Only data obtained with VLA's A-configuration is suitable
for combination with eMERLIN data, because the largest angular scale to which eMERLIN
is sensitive at L-band is $2^{\prime\prime}$. Finally, confusion is not expected to be
an issue in A-configuration at the flux densities considered in this white paper
\citep{2012ApJ...758...23C}. Confusion will be an issue for any MeerKAT surveys
deeper than MIGHTEE tier-2 until the longest baselines are extended from 8 to 20~km.

\vspace{5mm}\noindent{\bf \large Choice of target field}

\vspace{1mm}\noindent If a deep option is selected for a VLASS, it will be critical to
perform the survey over a field with extensive multiwavelength data in order to maximize
scientific impact. Community involvement will be required to identify which of the
existing multiwavelength extragalactic survey fields is best suited to a heavy
investment of VLA time. Some points to consider:
\begin{itemize}

\item	GOODS-N (12:36+62), XMM-LSS (2:25-5), COSMOS (10:00+2), the Lockman Hole (10:53+57),
	and the Groth Strip (14:17+52) are the most intensively observed fields in the
	Northern sky.

\item	The community may be interested in choosing a field visible to Southern facilities
	such as the Cerro Chajnantor Atacama Telescope (CCAT) and the Atacama Large
	Millimeter Array (ALMA), for example.

\item	Of the four 10~deg$^2$ deep drilling fields identified by the Large Synoptic Survey
	Telescope (LSST) Science Council, two are suitable for a deep VLA survey in
	A-configuration: XMM-LSS and COSMOS. The former is also one of the Dark Energy
	Survey (DES) deep supernova survey fields.

\item	MeerKAT-MIGHTEE tier-2 will target the XMM-LSS and COSMOS fields. This may present
	an opportunity to combine MeerKAT and VLA data to increase sensitivity to
	extended structures.

\item	MeerKAT LADUMA/MIGHTEE tier-3 will target the CDF-S (3:32-28), while eMERLIN-eMERGE
	tier-1 will target the HDF-N. A deep VLASS on a different field will bring the number
	of deep fields to three, enabling more robust scientific conclusions to be drawn in
	the lead-up to the SKA than if only one or two fields were observed.

\item	The VLA-CHILES survey is currently imaging a single pointing at L-band
	in B-configuration to a continuum sensitivity of 0.7~$\mu$Jy in the COSMOS field;
	if a deep VLASS is pursued in the same field, the CHILES data can be included.

\end{itemize}

\vspace{5mm}\noindent{\bf \large Deep field science at L-band}

\vspace{1mm}\noindent A brief selection of science topics that can be addressed with a
deep L-band survey in A-configuration is presented below. While a number of these topics
will also be addressed within the next few years by facilities such as WSRT, MeerKAT,
and ASKAP, many demand the greater spatial resolving power of the VLA. As a result, a
deep L-band VLASS will be of lasting value to the astronomical community well into the SKA-era. 
\begin{itemize}

\item	Faint radio galaxy populations; morphologies, spectral indices, and environments
	
\item	Evolution of supermassive black holes, active galaxies (particularly low power),
	and star formation across cosmic time; luminosity functions

\item	Strong gravitational lenses, high redshift radio galaxies

\item	Much like the VLA-CHILES survey, the flexible WIDAR correlator could be set up for a deep
	VLASS to provide fine spectral resolution over certain redshift ranges\linebreak for HI
	absorption science, and possibly HI emission and recombination line science

\item	Dark energy and dark matter using radio weak lensing \citep{2002ApJ...570L..51B,2006ApJ...650L..21M,
	2011MNRAS.410.2057B,2011ApJ...735L..23B}; high spatial resolution radio polarization and HI
	data will be very valuable
	
\item	The evolution of magnetism in galaxies and large scale structure using radio polarization
	data \citep[e.g.][]{2013ApJ...767..150A}; commensal P-band observations could be obtained with
	L-band, providing improved resolution in Faraday space

\item	Transient and variable sources

\item	Degenerate and non-degenerate radio stars

\item	Develop a gold standard multiwavelength catalog for machine learning algorithms
	to catalog much wider-area and shallower surveys; for example, SKA1-survey

\end{itemize}

\renewcommand{\refname}{\bf \large References}

\begin{multicols}{2}

\end{multicols}

\newpage

\begin{figure}
 \centerline{\includegraphics[clip, angle=-90, totalheight=130mm]{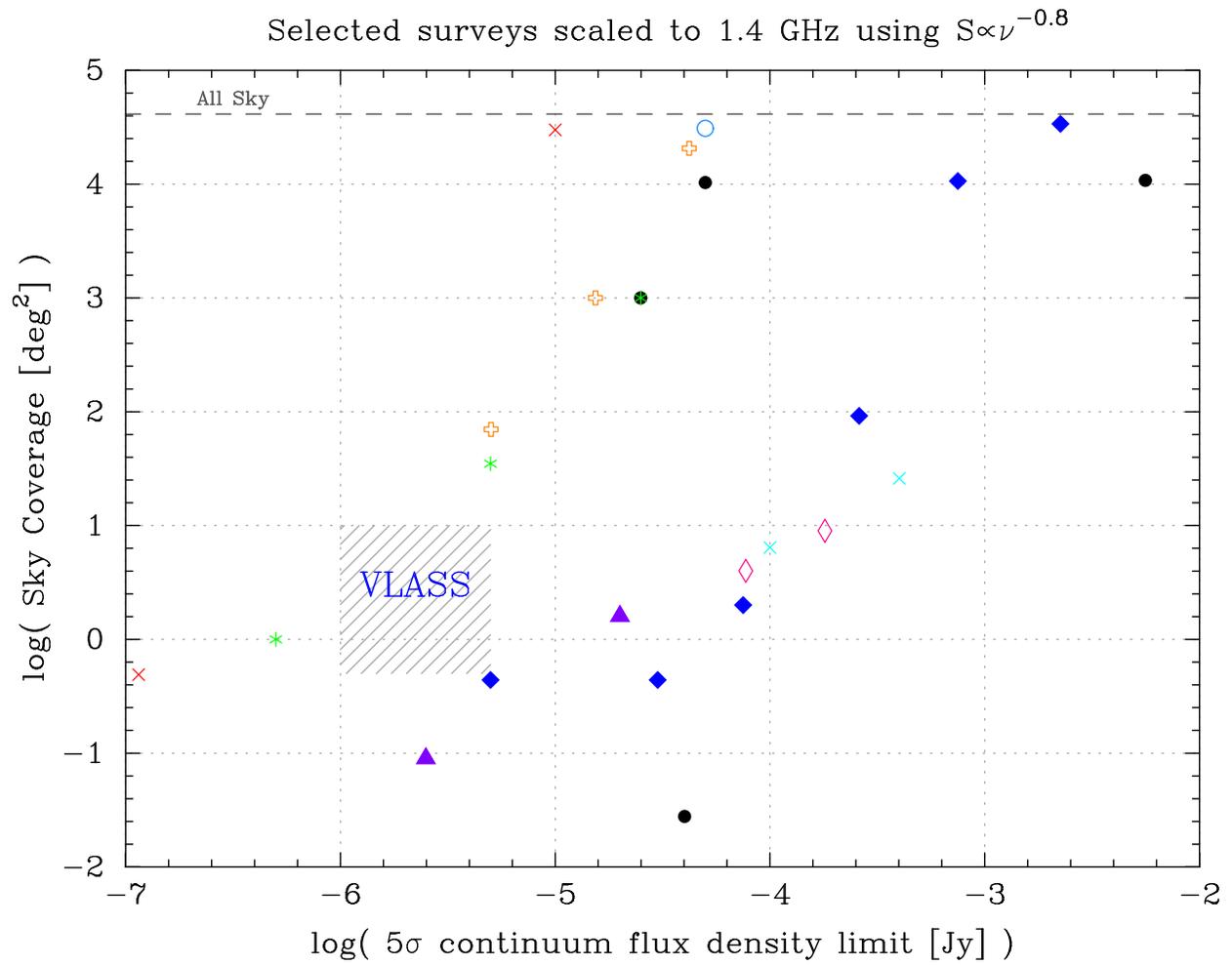}}\vspace{5mm}
 \caption{Reproduction of Fig.~\ref{fig:surveys} with labels removed for ease of viewing.}
 \label{fig:surveys2}
\end{figure}

\end{document}